\documentclass[10pt,a4paper]{article}
\usepackage[a4paper, total={6.5in, 9in}]{geometry}
\usepackage{amsmath}
\usepackage{amsfonts}
\usepackage{amssymb}
\usepackage{graphicx, float}
\usepackage{authblk}

\title{Evaluation of effective medium theories for spherical nano-shells}
\author[1]{Yael Gutierrez}
\author[1]{Dolores Ortiz}
\author[1]{Rodrigo Alcaraz de la Osa}
\author[1,2]{Juan M. Sanz}
\author[1]{Jose M. Saiz}
\author[1]{Francisco Gonzalez}
\author[1,*]{Fernando Moreno}

\affil[1]{Group of Optics, Department of Applied Physics, University of Cantabria, Faculty of Sciences, Avda. Los Castros s/n 39005 Santander, Spain.}
\affil[2]{R\&D Department, Textil Santanderina, S.A., Avda. Textil Santanderina, s/n 39500 Cabezon de la Sal, Spain}
\date{}

\begin{document}

\maketitle

\begin{abstract}
	Current effective medium theories for nano-shells are reviewed. A new method for calculating the effective dielectric function of a core-shell nanoparticle is presented and compared with existing theories showing clear advantages in most conditions. It consists of introducing radiating effects in the polarizability of the effective sphere, and considering the exact polarizability of the core-shell constructed from the Mie scattering coefficient. This new approach can be considered as an useful tool for designing coated particles with desired plasmonic properties and engineering the effective permittivity of composites with core-shell type inclusions which are used in photocatalysis and solar energy harvesting applications.
\end{abstract}

\section{Introduction}

Plasmonics is an active branch of Nanophotonics which studies the distribution of the electromagnetic field, and its local charge resonances (Localized Surface Plasmon Resonances, LSPRs) in subwavelength metallic nanostructures when electromagnetically irradiated \cite{Meier2007}. New advances in plasmonics require far more complex geometries made by combining different materials and shapes. One of the most widely used of these geometries is the core-shell spherical nanoparticle (NP), which proved to be a useful structure in a variety of fields, i.e. catalysis, biology, materials chemistry and sensors \cite{Gawande2015a}. 

There is a great interest in extending nanoplasmonics to the UV-range due its current and potential applications in, for instance, catalysis, biology and semiconductor technology \cite{Zhang2013a, Anker2008, Chowdhury2009,Taguchi2009}. Sanz et al. \cite{Sanz2013} studied several metals in order to find those whose properties made them more promising for UV-plasmonics. Two of the most compelling metals for this purpose are aluminum (Al) \cite{Knight2012, Knight2014} and magnesium (Mg) \cite{Sterl2015}. Interestingly, NPs made of these two metals suffer from the formation of a native oxide layer whose thickness depends on both the metal and the environmental exposure conditions \cite{Rai2006, Sterl2015}. Very recently it was shown that oxidation control may be even used for tuning the resonance of NPs \cite{Sterl2015}. Although, theoretically, these oxidized NPs can be modeled as metal-oxide core-shell NPs \cite{Gutierrez2016}, some authors have proposed to treat this type of nanostructures with effective medium theories \cite{Chettiar2012, Diaz-H.R2016}. For example, Knight et al. \cite{Knight2014} used the Bruggeman effective medium theory to describe the behavour of Al/Al$_{2}$O$_{3}$ nanodisks, and Kuzma et al. \cite{Kuzma2012a} employed the Maxwell-Garnet theory to model oxidized silver NPs. 

As an alternative, the coating of Al and Mg NPs with a layer of Gallium has been proposed  \cite{Wu2011, Wu2011a}, as a way to prevent the oxidation process. Gallium, a metal with plasmonic resonances in the UV of his own, forms a self-terminating layer of oxide of a few nanometers that prevents further oxidation \cite{Wu2007, Yang2014a}. This metal-metal nanostructures can be considered alloys \cite{Yang2014a}, and be modelled with effective medium theories \cite{DelaOsa2013}.

Effective medium theories constitute an approach for simulating the electromagnetic behavior of systems with embedded NPs in a matrix by obtaining an effective dielectric function of the ensemble. Usually these kind of approaches, like the one proposed by Maxwell-Garnett, require the dielectric function of both the inclusions and the matrix. However, if core-shell NPs are considered as inclusions, an effective dielectric functions of this type of particles is necessary to obtain the total effective dielectric function of the system. This procedure can be useful to predict the reflectance, transmittance and absorptance spectra of core-shell colloids and nanocomposites which are widely used in photocatalysis \cite{Das2015} or solar energy harvesting \cite{Zhang2013b, Pathak2016}.

In this work we have performed a detailed study of the current effective medium theories for spherical nano-shells (section \ref{sec:EMT}). A new procedure for obtaining the effective dielectric function of core-shell spherical nanoparticles is presented in section \ref{sec:NewApproach}. In section \ref{sec:Results}, the extinction efficiencies obtained from the effective dielectric functions are compared with the exact solution given by Mie theory in order to determine which one better describes the electromagnetic behaviour of a nano-shell. These comparisons will be done over some particular core-shell metal-oxide nano-shells, namely, Mg/MgO and Al/Al$_2$O$_3$ (subsection \ref{sec:Results}.\ref{ssec: DM}), and metal-metal NPs, Mg/Ga and Al/Ga (subsection \ref{sec:Results}.\ref{ssec: MM}). Finally, in section \ref{sec:Conclusions} the main conclusions are presented.

\section{Effective Medium Theories}\label{sec:EMT}
\begin{figure}[H]\centering
	\includegraphics[scale=0.40 ]{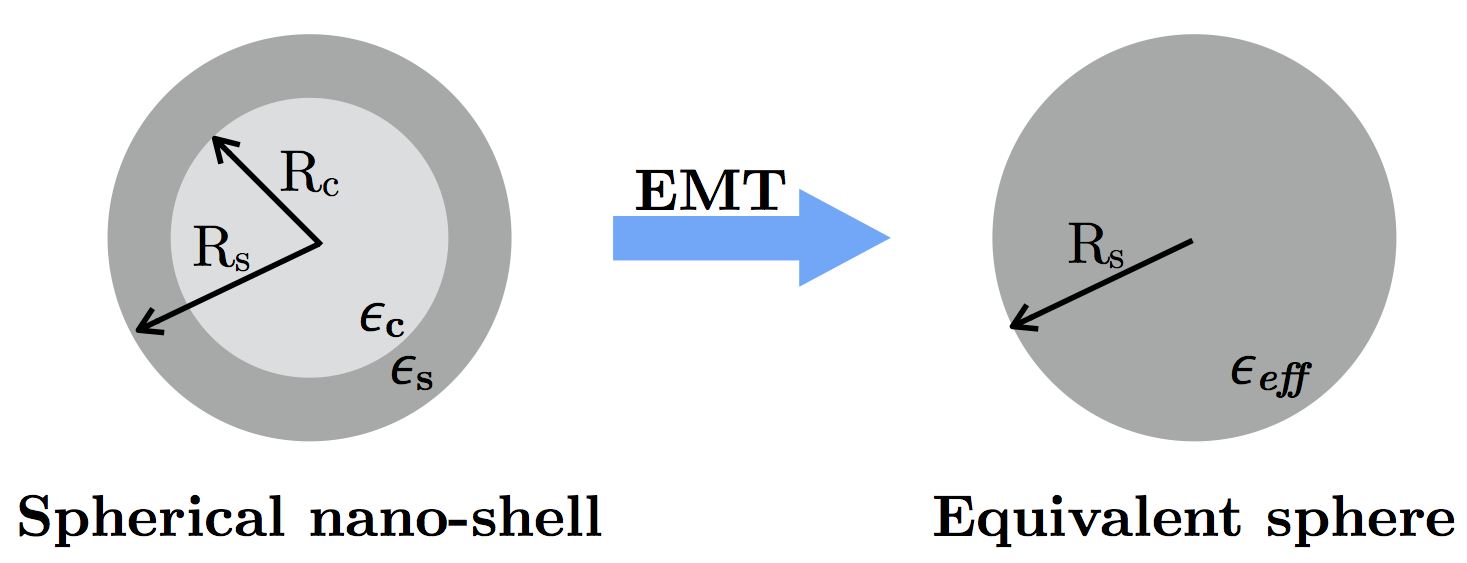}   
	\caption{Scheme of the effective medium problem for a spherical nano-shell.}\label{fig:scheme}
\end{figure}

Although the electromagnetic response of spherical core-shell NPs has analytical solution given by Mie Theory \cite{Bohren1998}, the effective medium theories (EMTs) give an alternative for dealing with this type of NPs. These EMTs allow us to obtain the dielectric function of an equivalent sphere with the same size and electromagnetic behaviour as a given nano-shell (Fig. \ref{fig:scheme}).

Here we review some of the existing EMTs applied to spherical core-shell NPs. 

\subsection{Weighted Average (WA)}
One of the simplest approaches to get the effective dielectric function ($\epsilon_{eff}$) of a core-shell nanoparticle is given by the weighted average of the dielectric functions \cite{ Kuzma2012a} of both the core ($\epsilon_{c}$) and the shell ($\epsilon_{s}$),
\begin{equation}\label{eq:WA}
\epsilon_{eff} = f\epsilon_{c} + (1-f)\epsilon_{s}
\end{equation}
where $f=R_c^3/R_s^3$ is the fraction of volume that occupies the core. When $f=1$ (core-material particle) $\epsilon_{eff} = \epsilon_{c}$. However, when $f=0$ (shell-material particle) $\epsilon_{eff} = \epsilon_{s}$.

\subsection{Bruggeman}
Bruggeman's effective medium theory \cite{Bruggeman1935, Bohren1998} is proposed for non-homegenous media in which two materials are present, the main one ---the matrix--- hosting the secondary one ---the inclusion. Both materials, however, are given the same importance, i.e., it is invariant to an interchange between the host matrix and the inclusions. In this approximation, the effective dielectric function $\epsilon_{eff}$ is given by 

\begin{equation}\label{eq:Bruggeman}
f\frac{\epsilon_{c} - \epsilon_{eff}}{\epsilon_{c} + 2\epsilon_{eff}} + (1-f)\frac{\epsilon_{s} - \epsilon_{eff}}{\epsilon_{s} + 2\epsilon_{eff}} = 0
\end{equation}

where $\epsilon_{s}$ and $\epsilon_{c}$ are the dielectric functions of the host (in our case the shell) and the inclusions (the core) respectively, and $f$ is the filling factor of the inclusions. If in \eqref{eq:Bruggeman}, we now consider the filling fraction of the host ($f'$) instead of the one of the inclusion ($f'=(1-f)$), we obtain an equivalent expression. This means that which component is labelled as inclusion or host does not matter, both are interchangeable.

This method was used by Knight et al. \cite{Knight2014} to model the behaviour of oxidized Al nanodisks. 

\subsection{Maxwell-Garnet (MG)}
The Maxwell-Garnet theory \cite{Bohren1998, Garnett} considers the case of inclusions randomly spread in a continuous matrix. This theory is based on the Clausius-Mossotti relation, which relates the polarizability ($\alpha$) with the dielectric function ($\epsilon$). The effective dielectric function of a suspension of small spheres in a host material is given by

\begin{equation}\label{eq:MG_initial}
\frac{\epsilon_{eff}-\epsilon_{h}}{\epsilon_{eff}+2\epsilon_{h}} = f
\frac{\epsilon_{I}-\epsilon_{h}}{\epsilon_{I}+2\epsilon_{h}}
\end{equation}

where $\epsilon_{h}$ and $\epsilon_{I}$ are the dielectric functions of host and inclusions materials, and $f$ is the volume fraction of the embedded particles.
Because a core-shell nanoparticle can be considered as formed by an inclusion (core with dielectric function $\epsilon_{c}$) in a host medium (shell with dielectric function $\epsilon_{s}$) where $\epsilon_{I} = \epsilon_{c}$ and $\epsilon_{h} = \epsilon_{s}$, its effective dielectric function can be expressed as

\begin{equation}\label{eq:MG-simply}
\epsilon_{eff} = \epsilon_{s}\frac{(\epsilon_{c} + 2\epsilon_{s})+2f(\epsilon_{c} - \epsilon_{s})}
{(\epsilon_{c} + 2\epsilon_{s})-f(\epsilon_{c} - \epsilon_{s})}
\end{equation}

Again, $f$ is the filling factor that in the case of a spherical nano-shell is given by $f=R_c^3/R_s^3$. Notice that when $f=1$ (core-material particle) $\epsilon_{eff} = \epsilon_{c}$. On the contrary, when $f=0$ (shell-material particle) $\epsilon_{eff} = \epsilon_{s}$.  

\subsection{Internal Homogenization (IH)}
A more complex approach was introduced by Chettiar and Enghetta \cite{Chettiar2012} through the concept of internal homogenization. Through this process the polarizability of the equivalent sphere is equated to that of a core-shell in the electrostatic approximation	\cite{Bohren1998}. According to this, the effective dielectric function for a core-shell NP in vacuum is given by
\begin{equation}\label{eq:MG}
\frac{\epsilon_{eff}-1}{\epsilon_{eff}+2} = 
\frac{(\epsilon_{s}-1)(\epsilon_{c}+2\epsilon_{s}) + f(\epsilon_{c}-\epsilon_{s})(1+2\epsilon_{s})}
{(\epsilon_{s}+2)(\epsilon_{c}+2\epsilon_{s}) + 2f(\epsilon_{s}-1)(\epsilon_{c}-\epsilon_{s})}
\end{equation}where $f$ is the previously defined filling factor. 

By rearranging \eqref{eq:MG}, the effective dielectric function can be expressed as the one given by a MG EMT (\eqref{eq:MG-simply}) assuming a core embedded in a shell material medium with a filling factor of $f$. This equation, that is independent of the particle size, is only valid in the regime in which the size of the particle is much smaller than the incident wavelength (electrostatic approximation).

\subsection{Mie theory based Maxwell Garnet (MMG)}

A size dependent extension of the MG model (\eqref{eq:MG_initial}) was proposed by Doyle \cite{Doyle1989, Ruppin2000} for a suspension of metallic spheres. In this case, the polarizability of the core embedded in shell material is given in terms of the Mie coefficient $a_1$ \cite{Bohren1998} rather than the Clausius-Mossoti relation,
\begin{equation}\label{eq:MMG}
\frac{\epsilon_{eff}-\epsilon_{s}}{\epsilon_{eff}+2\epsilon_{s}} =f  \frac{3i}{2x^3}a_1
\end{equation}

The main problem of this approach is that, although it considers the size effects on the polarizability of the core-shell nanoparticle, no size effect is taken into account in the equivalent sphere polarizability. In addition, by considering the polarizability of the core embedded in shell material, the validity of this model is restricted to the limit $R_c<<R_s$.

\section{New Approach}\label{sec:NewApproach}

All the previous EMTs are either limited by the size of the particles (electrostatic approximation) or by the condition $R_c<<R_s$. In this work, we present a new model aimed to overcome these limitations. On the one hand, we will consider the exact polarizability of the core-shell nanoparticle ($\alpha^{cs}$) in terms of its Mie dipolar electric coefficient $a_1^{cs}$ \cite{Mulholland1994},
\begin{equation}\label{eq:alfa_cs}
\alpha^{cs} = \frac{3i}{2x^3}a_1^{cs}
\end{equation}
By doing this, size effects are being considered and the $R_c<<R_s$ condition is removed. 

On the other hand, size effects in the effective dielectric function of the equivalent sphere can be taken into account by introducing dynamic depolarization factors \cite{Meier1983}.
\begin{equation}\label{eq:MW}
\alpha = \frac{\epsilon-1}{(\epsilon+2)-(\epsilon-1)x^2-(\epsilon-1)(2i/3)x^3}
\end{equation}
where $x$ is the size paramenter given by $x = 2\pi R/ \lambda$ being $R$ the size of the particle $R = R_s$. 

So by considering the exact polarizability of the core-shell (\eqref{eq:alfa_cs}) and the depolarization factor in the polarizability of the equivalent sphere (\eqref{eq:MW}), we can express the effective dielectric function of an spherical nano-shell as 

\begin{equation}\label{eq:ExtIH}
\frac{\epsilon_{eff}-1}{(\epsilon_{eff}+2)-(\epsilon_{eff}-1)x^2-(\epsilon_{eff}-1)(2i/3)x^3} = \frac{3i}{2x^3}a_1^{cs}
\end{equation}

This approach is more of an internal homogenization process \cite{Chettiar2012} than a Maxwell-Garnet effective medium theory \cite{Doyle1989}. For this reason, from now on we will refer to this theory as the Extended Internal Homogenization (Ext. IH).

\section{Results} \label{sec:Results}

\begin{figure}[H]\centering
	\includegraphics[scale=0.29]{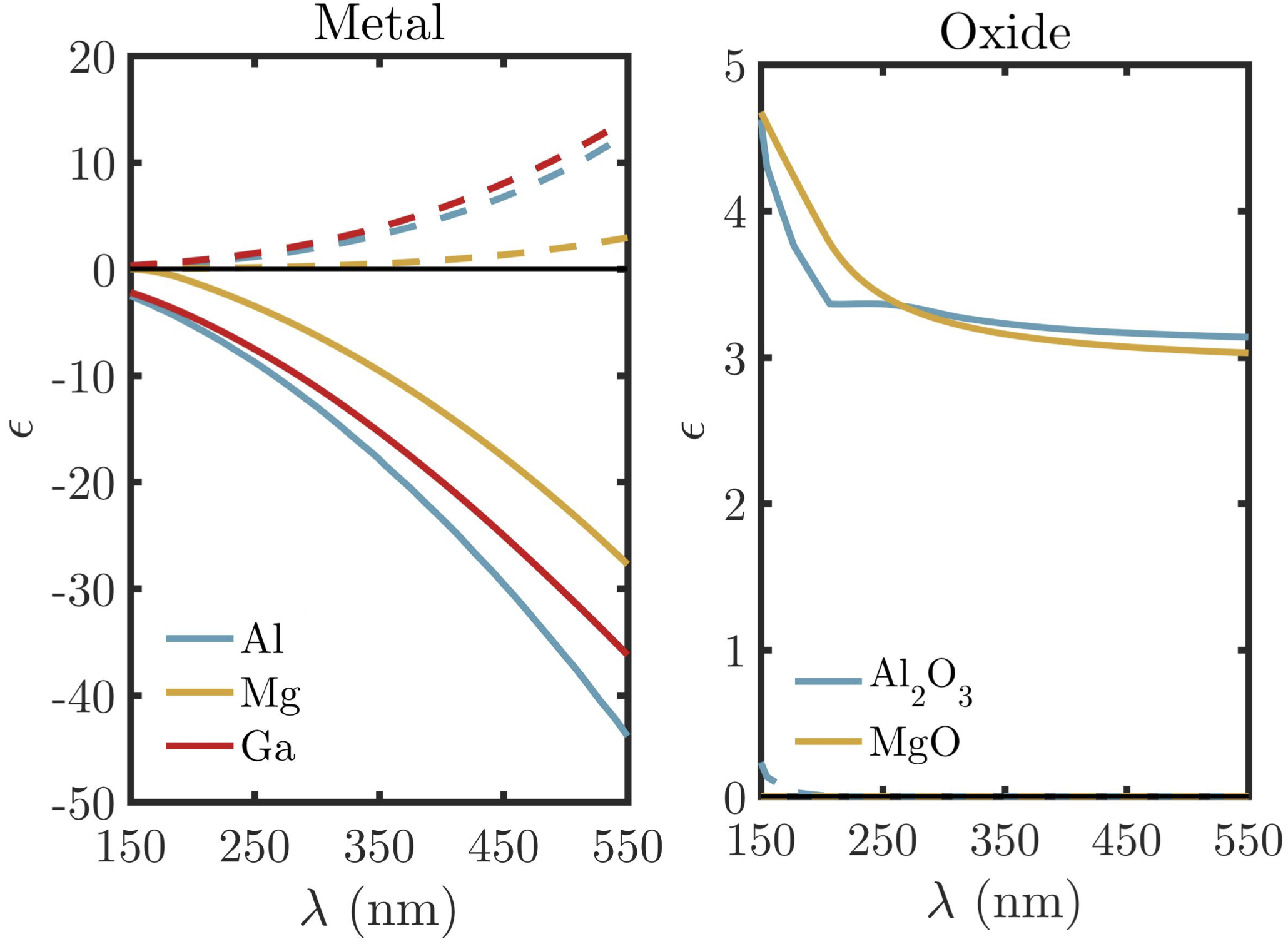}\caption{ Real (solid line, $\epsilon_r$) and imaginary (dashed line, $\epsilon_i$) part of the dielectric function for the metals Al, Mg, and Ga (left) and their  respective oxides Al$_2$O$_3$ and MgO (right).}\label{fig:epsilon}
\end{figure}

For evaluating the validity of each model we will compare their results with the exact solution given by Mie theory for the core-shell case. We will consider both dielectric/metal and metal/metal core-shell NPs. When considering dielectric/metal nanoparticles we will analyze oxide/metal nano-shells, namely Mg/MgO and Al/Al$_2$O$_3$. For the case of metal/metal core-shells we will study Mg/Ga and Al/Ga NPs. The real and imaginary parts of the dielectric function ($\epsilon = \epsilon_r + i \epsilon_i$) of these materials \cite{Palik1998, Losurdo2016} can be seen in Fig. \ref{fig:epsilon}. 

\subsection{Dielectric/Metal Core-Shell NP} \label{ssec: DM}

\begin{figure}[H]\centering
	\includegraphics[scale=0.60]{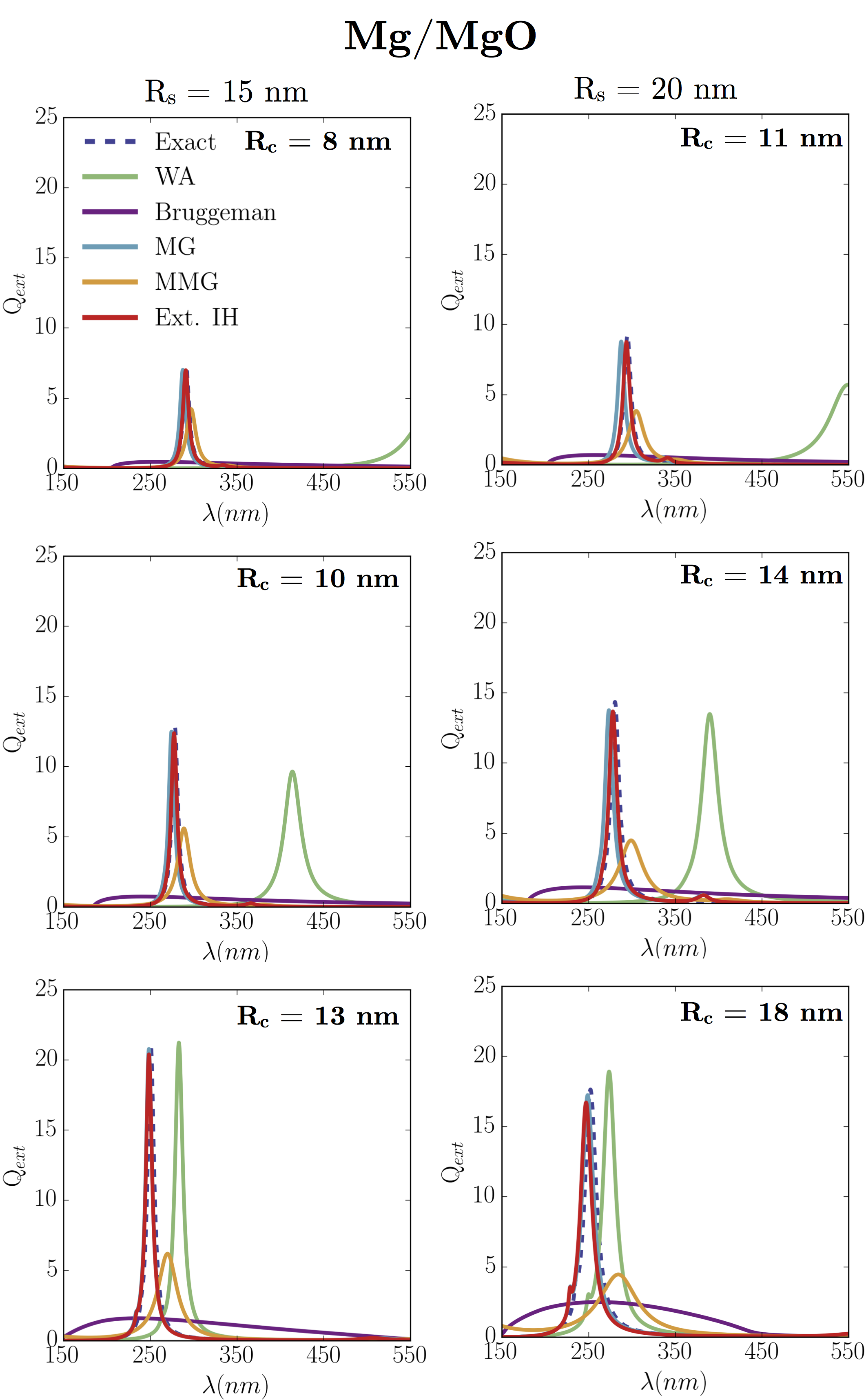}   
	\caption{Extinction efficiency ($Q_{ext}$) of Mg/MgO nano-shells of external radius $R_s = 15$ (left column) and $20$ nm (right column), and the different core sizes ($R_c$) indicated in the inset. Mie theory is compared with effective medium theories.}\label{fig:Mg_spectrum}
\end{figure}

\begin{figure}[H]\centering
	\includegraphics[scale=0.60]{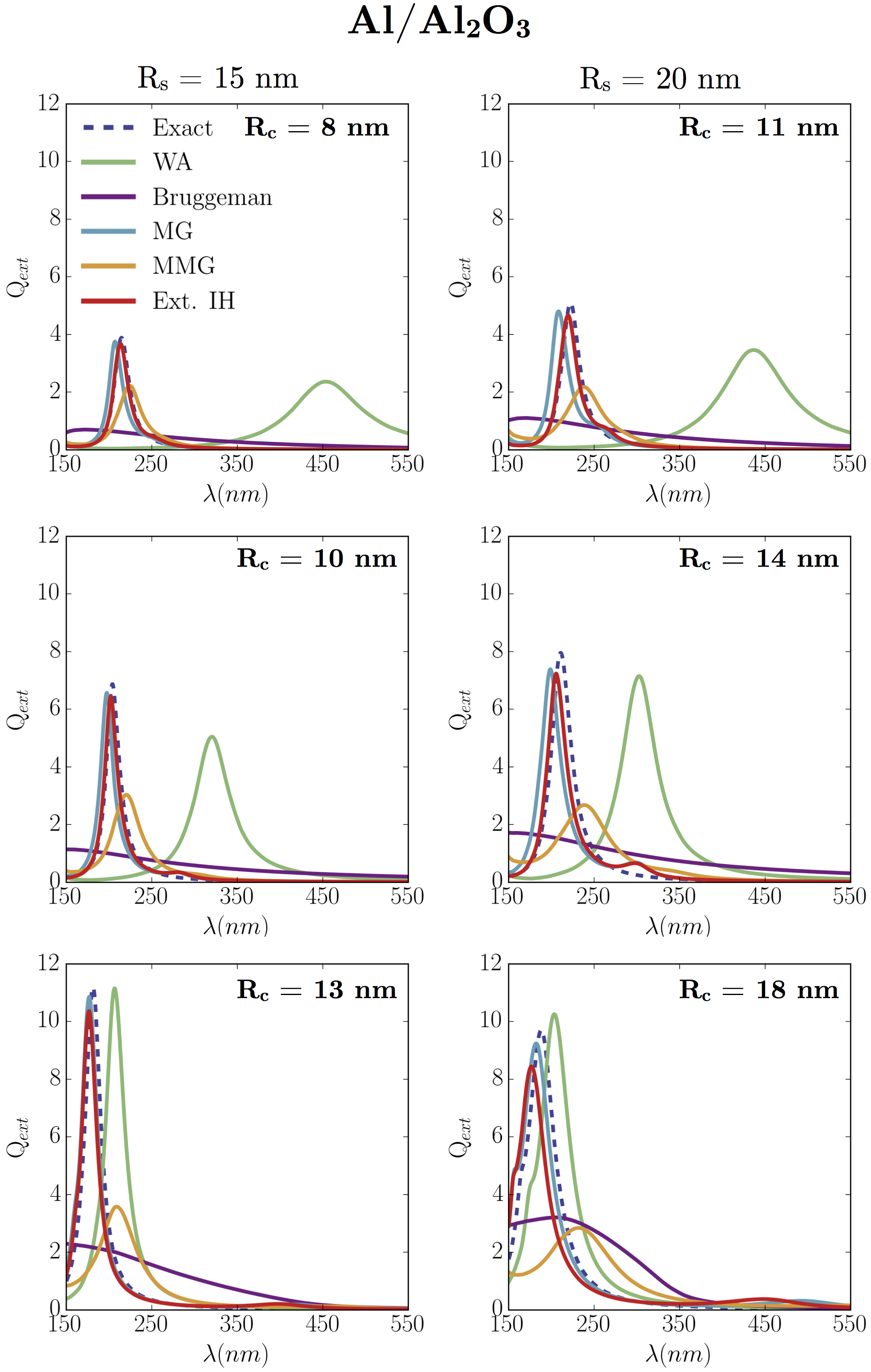}   
	\caption{Extinction efficiency ($Q_{ext}$) of Al/Al$_{2}$O$_{3}$ nano-shells of external radius $R_s = 15$ (left column) and $20$ nm (right column), and the different core sizes ($R_c$) indicated in the inset. Mie theory is compared with effective medium theories.}\label{fig:Al_spectrum}
\end{figure}

The extinction efficiency (Q$_{ext}$) of Mg/MgO and Al/Al$_{2}$O$_{3}$ nano-shells of external radius $R_{s} = 15$ nm and $R_{s} = 20$ nm is shown in Figs. \ref{fig:Mg_spectrum} and \ref{fig:Al_spectrum} as a function of the core size $R_c$. Mie theory (Exact) is compared with effective medium theories: the weigthed avarage (WA, \eqref{eq:WA}), the Bruggeman theory (\eqref{eq:Bruggeman}), the Maxwell-Garnet model (MG, \eqref{eq:MG-simply}), the Mie theory based extension introduced by Doyle (MMG,  \eqref{eq:MMG}), and finally, the extension of the internal homogenization model that we propose in this work (Ext IH, \eqref{eq:ExtIH}).

The studied range of $R_c$ goes from values of the oxide-shell width ($R_s - R_c$) of 2 nm to approximately half the value of $R_s$. Al forms an oxide layer that stops growing after a few nanometers preventing from further oxidation \cite{Rai2006}. However, the Mg oxidation process is more aggressive: the oxygen diffuses through the porous oxide leading to the complete destruction of the plasmonic response unless this procedure is performed under controlled conditions \cite{Sterl2015}. In all cases, oxide layers of a few nanometres are form. Taking into account the high influence of the ambient condition, the chosen values $R_c$ are realistic to describe the oxidized nanoparticles made of these two metals.

It is possible to see in Figs. \ref{fig:Mg_spectrum} and \ref{fig:Al_spectrum} how the Maxwell-Garnett theory (MG) and the proposed method (Ext. IH) give very similar results, which are the ones that better reproduces the exact calculations performed through Mie theory. On the contrary, the Bruggeman effective medium theory is the one that worst describes the plasmonic response of the nano-shells. The reason is that in this type of geometry we cannot assume that shell and core are interchangeable.

It also can be seen how the effective dielectric constant obtained through a weighted average (WA) does not reproduce the electromagnetic behaviour of the core-shell nanoparticles accurately, specially for the smallest values of $R_c$. Due to the construction of the expression (see \eqref{eq:WA}), for the largest values of the core radius ($Rc \approx Rs$), the response of the effective sphere is more similar to the one of the core-shell because $\epsilon_e \approx \epsilon_c$.

In the case of the Mie theory based Maxwell-Garnet (MMG), although it presents a good agreement at small values of $R_c$, it fails to reproduce the values of $Q_{ext}$. All these problems arise from the basis of this model, which considers the core embedded in shell material. When the metallic core is small, the evanescent field remains attached to it and dissipates within the dielectric shell, so we can consider the core to be embedded in shell material. However, as the shell thickness decreases the part of the evanescent field reaches beyond the surface of the outer shell \cite{Gutierrez2016}. It is in this regime where MMG fails because the core cannot be considered to be hosted in shell material.

\begin{figure}[H]\centering
	\includegraphics[scale=0.55]{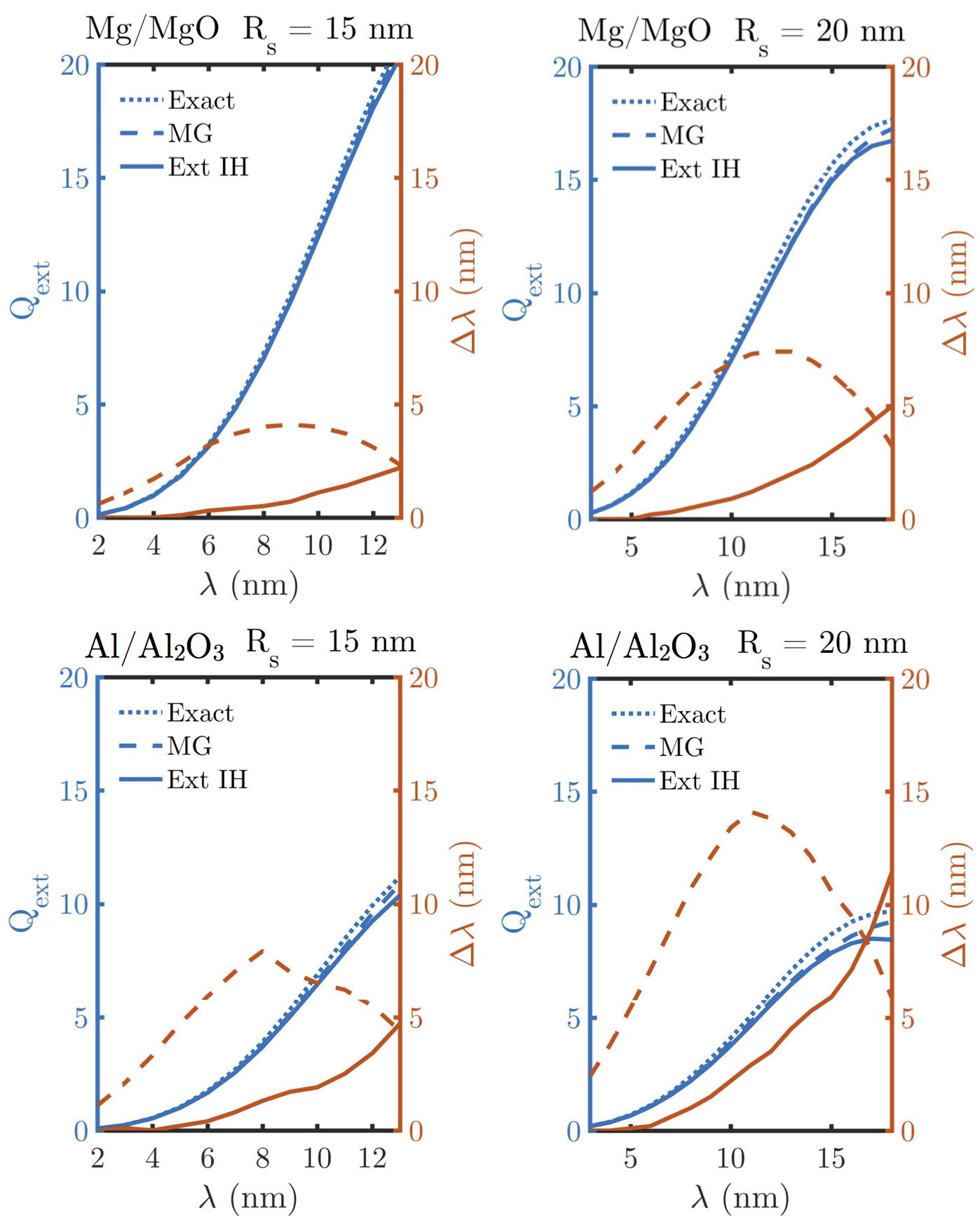}   
	\caption{$Q_{ext}$ (blue left axis) and spectral shift of the $Q_{ext}$ resonance peak (red right axis) predicted by MG and Ext. IH theories for a Mg/MgO (top row) and Al/Al$_{2}$O$_{3}$ (bottom row) core-shell nanoparticles of size $R_s=15$ (left column) and $R_s=20$nm (right column) with respect to the exact solution given by Mie Theory as function of the core size $R_c$.}\label{fig:delta}
\end{figure}

Figure \ref{fig:delta} shows the spectral shift of the $Q_{ext}$ resonance peak and its value predicted by MG and Ext. IH theories for core-shell Mg/MgO and Al/Al$_{2}$O$_{3}$ NPs with respect to the exact solution given by Mie theory for different core sizes $R_c$. Once again the particle sizes have been chosen to be $R_s=15$ and $R_s=20$ nm. 

For the smaller nano-shell size, it can be seen how the Ext. IH model reproduces better the behaviour of the core-shell NPs than the MG model, specially for the smallest core radii. For the biggest values of $R_c$, the MG model gives a better prediction because in the limit in which $R_c = R_s$ the effective permittivity tends to the dielectric function of the core. This difference can be better seen in the case of the $R_s=20$ nm nano-shell. The reason of this difference is that the polarizability of the core-shell in this method is only related with the electric dipolar term $a_1$. For metals with plasmonic behaviour in the UV range, a size of 20 nm (size parameter $x\approx 0.5$) involves higher multipolar terms so the results are deviated from the exact solution. It can be seen that when $R_c>17$ nm, the MG model becomes closer to the exact solution because in the limit where $R_c \approx R_s$ the effective dielectric function takes the value of $\epsilon_{eff} \approx \epsilon_c$. However, for $R_c<17$ nm the Ext. IH model gives a better prediction.

The range of applicability of the proposed model is restricted to nanoshells with only dipolar response. Those behaviours caused by higher multipolar order will not be predicted by this new approach. Because this approach has been optimized to model metal-oxide core-shell nanoparticles, it is useful to model core-shell NP with a metallic core and dielectric shell with a dielectric function $\epsilon_s$ ranging from $1$ to $4$. Although for the thinner shells (thickness $< 2-3$ nm) the proposed effective medium theory gives a worse prediction than the Maxwell-Garnet theory, for thicker shells our approach reproduces better the exact solution.   

\subsection{Metal/Metal Core-Shell NP} \label{ssec: MM}

\begin{figure}[H]\centering
	\includegraphics[scale=0.52]{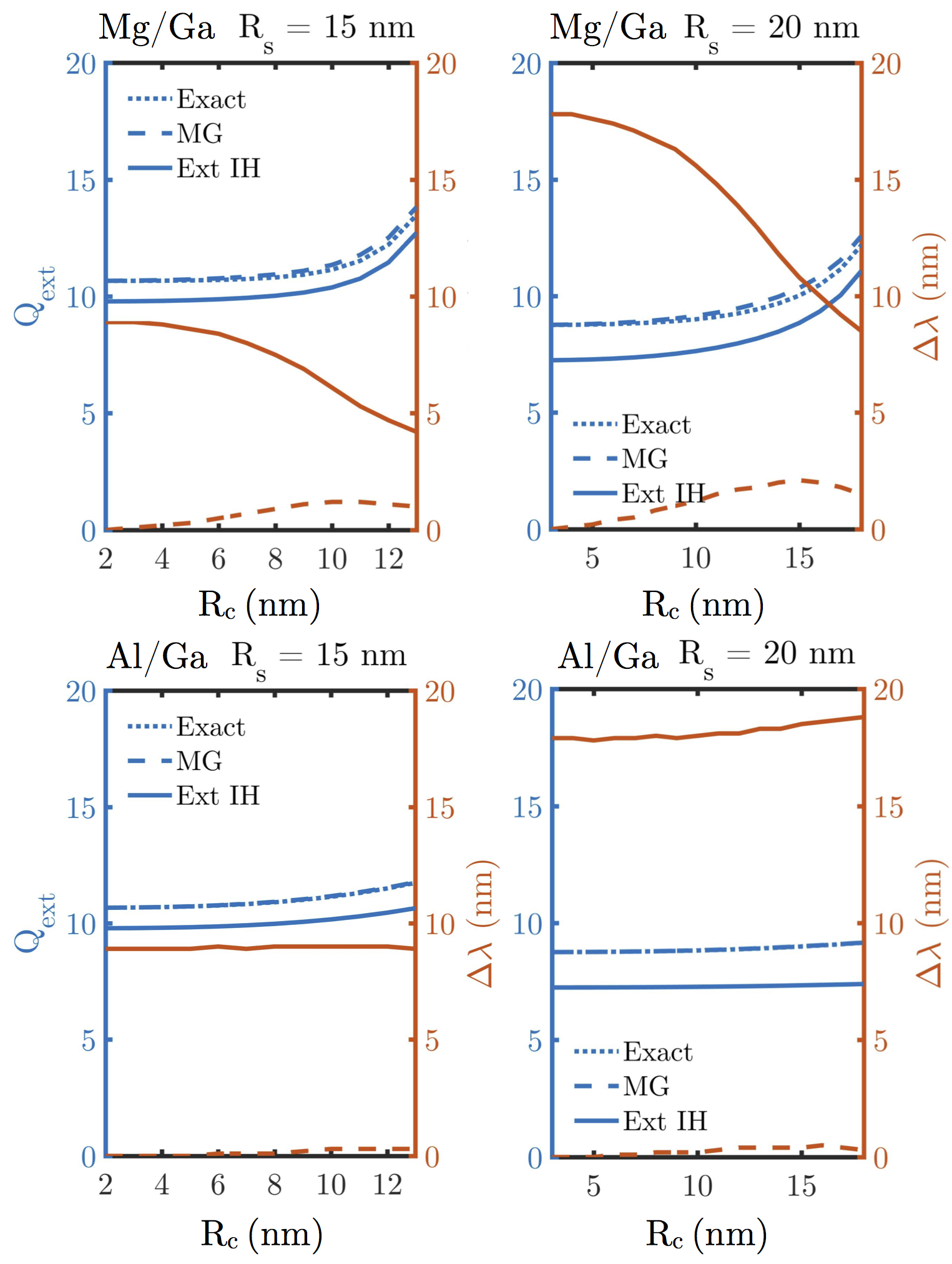}   
	\caption{$Q_{ext}$ (blue left axis) and spectral shift of the $Q_{ext}$ resonance peak (red right axis) predicted by MG and Ext. IH theories for a Mg/Ga (top row) and Al/Ga (bottom row) core-shell nanoparticles of size $R_s=15$ (left column) and $R_s=20$nm (right column) with respect to the exact solution given by Mie Theory as function of the core size $R_c$.}\label{fig:delta_Ga}
\end{figure}

Because the MG and the Ext. IH approaches were shown to better reproduce the behaviour of oxide-metal core-shells, here we analyze the validity of these two models when metal-metal core-shells are being considered. Figure \ref{fig:delta_Ga} shows the values of the $Q_{ext}$ resonance peak predicted by MG and Ext. IH. approaches for Mg/Ga and Al/Ga core-shell NP as the core size ($R_c$) increases. The spectral shift with respect to the exact solution given by Mie theory is also plotted. As in the previous case the particle sizes have been chosen to be $R_s=15$ and $R_s=20$ nm.

It can be seen how for the metal-metal core-shell nanoparticles the MG theory better reproduces the $Q_{ext}$ spectral behaviour of this kind of NPs more accurately than the Ext. IH approach. The Mg/Ga particle is the one that presents the bigger deviation from the exact behaviour. This can be understood attending to the MG formula construction and the Ga and Mg dielectric constants. The optical constant of Mg are quite different from those of Ga (see Fig. \ref{fig:epsilon}). The MG theory formula builds values of the optical constants of these type of NPs bounded between the dielectric function of the core and the shell. The similar the dielectric constants of the core and the shell, the smoother is the transition between the two regimes. For example, when we consider a dielectric-metallic core-shell, where the dielectric constants have very different behaviours, the prediction of the plasmonic behaviour is only good in the regions dominated by the shell ($R_c\approx 0$ or $f\approx 0$) or the core ($R_c\approx R_s$ or $f\approx 1$) (see \ref{ssec: DM}). In the case of Al/Ga core-shell, which has very similar dielectric constant, the predicted and the exact solution almost match.

The Ext. IH gives a poor approximation of the exact behavior of the metal-metal core-shell NP. When analyzing dielectric-metal core-shell, we observed that as the metallic core increases the predicted plasmonic response was deviated from the exact one. This was due to the appearance of higher multipolar orders. However, in the case of metal-metal core-shell NPs, the metallic particle keeps its size constant, so there is a continuous appearance of the quadrupolar mode. For the $R_s=15$ nm particle the deviation from the exact solution is smaller because the smaller the NP, the purer its dipolar response. Considering this fact, the metal-metal core-shell NPs of these sizes are out of the range of applicability of this approach.

\section{Conclusions} \label{sec:Conclusions}

In this work we have reviewed the current methods used to calculate the effective dielectric function of core-shell spherical NPs. A new approach to calculate this magnitude has been presented and compared with the existing models. Basically, it works by introducing introducing radiating effects in the polarizability of the effective sphere, and considering the exact polarizability of the core-shell constructed from the Mie scattering coefficient. This new method is valid outside the electrostatic approximation, one of the limitations of the current EMTs for this type of nanoparticles. This new approach has shown a better performance than the current EMTs for dielectric-metallic core-shell NPs. We believe that this method can be helpful for predicting the reflectance, transmittance, absorptance spectra, and engineer the effective permittivity of composites and colloids with core-shell type inclusions used in photocatalysis or solar energy harvesting.

\section{Funding Information}
This research has been supported by MICINN (Spanish Ministry of Science and Innovation, project FIS2013-45854-P) and sponsored by the Army Research Laboratory and was accomplished under Cooperative Agreement Number W911NF-17-2-0023. The views and conclusions contained in this document are those of the authors and should not be interpreted as representing the official policies, either expressed or implied, of the Army Research Laboratory or the U.S. Government. The U.S. Government is authorized to reproduce and distribute reprints for Government purposes notwithstanding any copyright notation herein. Y. G. wants to thank the University of Cantabria for her FPU grant.


\begin{thebibliography}{10}
	
	\bibitem{Meier2007}
	Stefan~Alexander Maier.
	\newblock {\em {Plasmonics: Fundamentals and Applications}}.
	\newblock Springer US, Boston, MA, 2007.
	
	\bibitem{Gawande2015a}
	Manoj~B. Gawande, Anandarup Goswami, Tewodros Asefa, Huizhang Guo, Ankush~V.
	Biradar, Dong-Liang Peng, Radek Zboril, and Rajender~S. Varma.
	\newblock {Core?shell nanoparticles: synthesis and applications in catalysis
		and electrocatalysis}.
	\newblock {\em Chem. Soc. Rev.}, 44(21):7540--7590, 2015.
	
	\bibitem{Zhang2013a}
	Xuming Zhang, Yu~Lim Chen, Ru-Shi Liu, and Din~Ping Tsai.
	\newblock {Plasmonic photocatalysis.}
	\newblock {\em Reports on progress in physics. Physical Society (Great
		Britain)}, 76(4):046401, 2013.
	
	\bibitem{Anker2008}
	Jeffrey~N Anker, W~Paige Hall, Olga Lyandres, Nilam~C Shah, Jing Zhao, and
	Richard~P {Van Duyne}.
	\newblock {Biosensing with plasmonic nanosensors.}
	\newblock {\em Nature materials}, 7(6):442--453, 2008.
	
	\bibitem{Chowdhury2009}
	Mustafa~H. Chowdhury, Krishanu Ray, Stephen~K. Gray, James Pond, and Joseph~R.
	Lakowicz.
	\newblock {Aluminum nanoparticles as substrates for metal-enhanced fluorescence
		in the ultraviolet for the label-free detection of biomolecules}.
	\newblock {\em Anal. Chem.}, 81(4):1397--1403, 2009.
	
	\bibitem{Taguchi2009}
	Atsushi Taguchi, Norihiko Hayazawa, Kentaro Furusawa, Hidekazu Ishitobi, and
	Satoshi Kawata.
	\newblock {Deep-UV tip-enhanced Raman scattering}.
	\newblock {\em Journal of Raman Spectroscopy}, 40(9):1324--1330, sep 2009.
	
	\bibitem{Sanz2013}
	J.~M. Sanz, D.~Ortiz, R.~{Alcaraz de la Osa}, J.~M. Saiz, F.~Gonz{\'{a}}lez,
	a.~S. Brown, M.~Losurdo, H.~O. Everitt, and F.~Moreno.
	\newblock {UV Plasmonic Behavior of Various Metal Nanoparticles in the Near-
		and Far-Field Regimes: Geometry and Substrate Effects}.
	\newblock {\em The Journal of Physical Chemistry C}, 117(38):19606--19615, sep
	2013.
	
	\bibitem{Knight2012}
	Mark~W Knight, Lifei Liu, Yumin Wang, Lisa Brown, Shaunak Mukherjee, Nicholas~S
	King, Henry~O Everitt, Peter Nordlander, and Naomi~J Halas.
	\newblock {Aluminum plasmonic nanoantennas.}
	\newblock {\em Nano letters}, 12(11):6000--6004, 2012.
	
	\bibitem{Knight2014}
	Mark~W. Knight, Nicholas~S. King, Lifei Liu, Henry~O. Everitt, Peter
	Nordlander, and Naomi~J. Halas.
	\newblock {Aluminum for plasmonics}.
	\newblock {\em ACS Nano}, 8(1):834--840, 2014.
	
	\bibitem{Sterl2015}
	Florian Sterl, Nikolai Strohfeldt, Ramon Walter, Ronald Griessen, Andreas
	Tittl, and Harald Giessen.
	\newblock {Magnesium as Novel Material for Active Plasmonics in the Visible
		Wavelength Range}.
	\newblock {\em Nano Letters}, 15(12):7949--7955, dec 2015.
	
	\bibitem{Rai2006}
	A.~Rai, K.~Park, L.~Zhou, and M.~R. Zachariah.
	\newblock {Understanding the mechanism of aluminium nanoparticle oxidation}.
	\newblock {\em Combustion Theory and Modelling}, 10(5):843--859, oct 2006.
	
	\bibitem{Gutierrez2016}
	Yael Gutierrez, Dolores Ortiz, Juan~M Sanz, Jose~M Saiz, Francisco Gonzalez,
	Henry~O Everitt, and Fernando Moreno.
	\newblock {How an oxide shell affects the ultraviolet plasmonic behavior of Ga,
		Mg, and Al nanostructures}.
	\newblock {\em Optics Express}, 24(18):20621, sep 2016.
	
	\bibitem{Chettiar2012}
	Uday~K. Chettiar and Nader Engheta.
	\newblock {Internal homogenization: Effective permittivity of a coated sphere}.
	\newblock {\em Optics Express}, 20(21):22976, 2012.
	
	\bibitem{Diaz-H.R2016}
	Rafael Diaz-H.R, Raul Esquivel-Sirvent, and Cecilia Noguez.
	\newblock {Plasmonic Response of Nested Nanoparticles with Arbitrary Geometry}.
	\newblock {\em The Journal of Physical Chemistry C}, 120(4):2349--2354, 2016.
	
	\bibitem{Kuzma2012a}
	Anton Kuzma, Martin Weis, Sona Flickyngerova, Jan Jakabovic, Alexander Satka,
	Edmund Dobrocka, Juraj Chlpik, Julius Cirak, Martin Donoval, Peter Telek,
	Frantisek Uherek, and Daniel Donoval.
	\newblock {Influence of surface oxidation on plasmon resonance in monolayer of
		gold and silver nanoparticles}.
	\newblock {\em Journal of Applied Physics}, 112(10):103531, 2012.
	
	\bibitem{Wu2011}
	Pae~C. Wu, Maria Losurdo, Tong~Ho Kim, Borja Garcia-Cueto, Fernando Moreno,
	Giovanni Bruno, and April~S. Brown.
	\newblock {Ga-Mg Core-shell nanosystem for a novel full color plasmonics}.
	\newblock {\em Journal of Physical Chemistry C}, 115(28):13571--13576, 2011.
	
	\bibitem{Wu2011a}
	Pae~C. Wu, Tong-Ho Kim, Alexandra Suvorova, Maria Giangregorio, Martin
	Saunders, Giovanni Bruno, April~S. Brown, and Maria Losurdo.
	\newblock {GaMg Alloy Nanoparticles for Broadly Tunable Plasmonics}.
	\newblock {\em Small}, 7(6):751--756, 2011.
	
	\bibitem{Wu2007}
	Pae~C Wu, Tong-Ho Kim, April~S. Brown, Maria Losurdo, Giovanni Bruno, and
	Henry~O. Everitt.
	\newblock {Real-time plasmon resonance tuning of liquid Ga nanoparticles by in
		situ spectroscopic ellipsometry}.
	\newblock {\em Applied Physics Letters}, 90(10):103119, 2007.
	
	\bibitem{Yang2014a}
	Yang Yang, Neset Akozbek, Tong-ho Kim, Juan~Marcos Sanz, Fernando Moreno, Maria
	Losurdo, April~S Brown, and Henry~O Everitt.
	\newblock {Ultraviolet?Visible Plasmonic Properties of Gallium Nanoparticles
		Investigated by Variable-Angle Spectroscopic and Mueller Matrix
		Ellipsometry}.
	\newblock {\em ACS Photonics}, 1(7):582--589, jul 2014.
	
	\bibitem{DelaOsa2013}
	R~{Alcaraz de la Osa}, F~Moreno, and J~M Saiz.
	\newblock {A new approach for modeling composite materials}.
	\newblock {\em Optics Communications}, 291:405--411, 2013.
	
	\bibitem{Das2015}
	Sourav Das, Sayantan Sinha, Mrutyunjay Suar, Soon~Il Yun, Amrita Mishra, and
	Suraj~K. Tripathy.
	\newblock {Solar-photocatalytic disinfection of Vibrio cholerae by using Ag@ZnO
		core-shell structure nanocomposites}.
	\newblock {\em Journal of Photochemistry and Photobiology B: Biology},
	142:68--76, 2015.
	
	\bibitem{Zhang2013b}
	Wei Zhang, Michael Saliba, Samuel~David Stranks, Yao Sun, Xian Shi, Ulrich
	Wiesner, and Henry~J. Snaith.
	\newblock {Enhancement of Perovskite-Based Solar Cells Employing Core?Shell
		Metal Nanoparticles}.
	\newblock {\em Nano Letters}, 13(9):4505--4510, sep 2013.
	
	\bibitem{Pathak2016}
	Nilesh~Kumar Pathak, Nikhil Chander, Vamsi~K. Komarala, and R.~P. Sharma.
	\newblock {Plasmonic Perovskite Solar Cells Utilizing Au@SiO2 Core-Shell
		Nanoparticles}.
	\newblock {\em Plasmonics}, pages 1--8, may 2016.
	
	\bibitem{Bohren1998}
	Craig~F. Bohren and Donald~R. Huffman, editors.
	\newblock {\em {Absorption and Scattering of Light by Small Particles}}.
	\newblock Wiley-VCH Verlag GmbH, Weinheim, Germany, apr 1998.
	
	\bibitem{Bruggeman1935}
	D.~A.~G. Bruggeman.
	\newblock {Berechnung verschiedener physikalischer Konstanten von heterogenen
		Substanzen. I. Dielektrizit{\"{a}}tskonstanten und Leitf{\"{a}}higkeiten der
		Mischk{\"{o}}rper aus isotropen Substanzen}.
	\newblock {\em Annalen der Physik}, 416(7):636--664, 1935.
	
	\bibitem{Garnett}
	J.~C. {Maxwell Garnett}.
	\newblock {Colours in Metal Glasses, in Metallic Films, and in Metallic
		Solutions. II}.
	\newblock {\em Philosophical Transactions of the Royal Society A: Mathematical,
		Physical and Engineering Sciences}, 205(387-401):237--288, may 1906.
	
	\bibitem{Doyle1989}
	William~T. Doyle.
	\newblock {Optical properties of a suspension of metal spheres}.
	\newblock {\em Physical Review B}, 39(14):9852--9858, may 1989.
	
	\bibitem{Ruppin2000}
	R.~Ruppin.
	\newblock {Evaluation of extended Maxwell-Garnett theories}.
	\newblock {\em Optics Communications}, 182(4-6):273--279, aug 2000.
	
	\bibitem{Mulholland1994}
	George~W Mulholland, Craig~F Bohren, and Kirk~a Fuller.
	\newblock {Light Scattering by Agglomerates: Coupled Electric and Magnetic
		Dipole Method}.
	\newblock {\em Langmuir}, 10(8):2533--2546, aug 1994.
	
	\bibitem{Meier1983}
	M~Meier and A~Wokaun.
	\newblock {Enhanced fields on large metal particles: dynamic depolarization.}
	\newblock {\em Optics letters}, 8(11):581--583, 1983.
	
	\bibitem{Palik1998}
	Edward~D. Palik.
	\newblock {\em {Handbook of Optical Constants of Solids}}.
	\newblock Academic Press, 1998.
	
	\bibitem{Losurdo2016}
	Maria Losurdo, Alexandra Suvorova, Sergey Rubanov, Kurt Hingerl, and April~S.
	Brown.
	\newblock {Thermally stable coexistence of liquid and solid phases in gallium
		nanoparticles}.
	\newblock {\em Nature Materials}, 15(July):995--1002, 2016.
	
\end{thebibliography}
\end{document}